\documentclass[10pt]{iopart}

\usepackage{graphicx}

\usepackage{iopams}
\usepackage{amsfonts}

\begin{document}

\title[Classical Dirac particle]
{Is it possible to describe an electron by the evolution of a single point?}

\author{Mart\'{\i}n Rivas}
\address{Theoretical Physics Department, The University of the Basque Country,(retired)\\ 
Bilbao, Spain}
\ead{martin.rivas@ehu.eus}

\begin{abstract}
The answer to the title-question is affirmative. The analysis of the geometry of continuous and differentiable curves in three-dimensional Euclidean space suggests that the point represents the location of the center of charge of the electron, satisfies a system of ordinary differential equations of fourth order, and moves at the speed of light.
The center of mass of the electron is a different point and will be determined by the evolution of the center of charge. It is the relative motion of the center of charge around the center of mass that gives rise to the spin and magnetic properties. The invariance of the mass and the absolute value of the spin for the center of mass observer imply that in the interaction of the electron with an external electromagnetic field the particle has to radiate.
The analysis of a Poincar\'e invariant interaction of two electrons implies that the only relevant parameter that characterizes the interaction in a natural system of units, is the fine structure constant. The fundamentals of General Relativity have to be revisited.
\end{abstract}

\section{Introduction}
\label{Introduction}
Let us consider the possibility of describing a classical spinning electron by the evolution of a single point ${\bi r}$ in three-dimensional Euclidean space. From the physical point of view, it represents a mechanical system of three degrees of freedom. If we achieve this, it means that all observables related to the structure of the electron, like energy, linear momentum, spin, magnetic moment, electric dipole moment, etc., can be expressed in terms of the point ${\bi r}$ and its subsequent time derivatives.
\subsection{Geometry}
The theory of continuous and differentiable curves in three-dimensional space was developed in 1847 and is known as the Frenet-Serret formalism. The points of the curve are parameterized by the arc length $s$ that represents the Euclidean distance along the curve of this point ${\bi r}(s)$ from some initial position ${\bi r}(0)$. The function ${\bi r}(s)$ and its subsequent derivatives with respect to the parameter $s$ are continuous and differentiable functions of this parameter. 

Along the curve, and in terms of the derivatives of the point ${\bi r}(s)$ up to the third order, it is defined
a system of three orthogonal dimensionless unit vectors attached to the point. They are called the {\it tangent} ${\bi t}$, the {\it normal} ${\bi n}$ and the {\it binormal} ${\bi b}$. The intrinsic characteristics of any curve are the two continuous and differentiable scalar functions $\kappa(s)$ and $\tau(s)$, known as the {\it curvature} and {\it torsion}, respectively, defined at every point ${\bi r}(s)$. The curvature is expressed in terms of the first two derivatives of the function ${\bi r}(s)$ and the torsion involves the derivatives up to the third order. The knowledge of the two continuous functions  $\kappa(s)$ and $\tau(s)$, for $s\in[0,L]$, together the boundary conditions ${\bi r}(0)$ and ${\bi t}(0)$, ${\bi n}(0)$ and ${\bi b}(0)$, completely determine the curve of length $L$, in three dimensional space. 

In three-dimensional Euclidean space, given any arbitrary but continuous and differentiable curve ${\bi r}(s)$, the derivatives $d{\bi r}(s)/ds$, $d^2{\bi r}(s)/ds^2$ and $d^3{\bi r}(s)/ds^3$, will be in general, three linearly independent vectors. The three Frenet-Serret unit vectors are written in terms of the derivatives up to the third order of point ${\bi r}$, as:
\begin{equation}
{\bi t}=\frac{d{\bi r}}{ds},\quad {\bi n}=\frac{1}{\kappa}\frac{d^2{\bi r}}{ds^2},\quad {\bi b}=\frac{\kappa}{\tau}\frac{d{\bi r}}{ds}-\frac{\dot{\kappa}}{\kappa^2\tau}\frac{d^2{\bi r}}{ds^2}+\frac{1}{\kappa\tau}\frac{d^3{\bi r}}{ds^3},
\label{univect}
\end{equation}
or conversely
\begin{equation}
\frac{d{\bi r}}{ds}={\bi t},\quad \frac{d^2{\bi r}}{ds^2}=\kappa{\bi n},\quad \frac{d^3{\bi r}}{ds^3}=-\kappa^2{\bi t}+\dot\kappa{\bi n}+\kappa\tau{\bi b},
\label{derivat}
\end{equation}
where
\begin{equation}
\kappa=\bigg|\frac{d^2{\bi r}}{ds^2}\bigg|,\qquad 
\tau=\frac{1}{\kappa^2}\left(\frac{d{\bi r}}{ds}\times\frac{d^2{\bi r}}{ds^2}\right)\cdot\frac{d^3{\bi r}}{ds^3},
\label{capatau}
\end{equation}
and have dimensions of length$^{-1}$.

The derivatives of the unit vectors with respect to $s$, are expressed as:
\begin{equation}
\dot{\bi t}=\bomega\times{\bi t}=\kappa{\bi n},\quad\dot{\bi n}=\bomega\times{\bi n}=-\kappa{\bi t}+\tau{\bi b},\quad\dot{\bi b}=\bomega\times{\bi b}=-\tau{\bi n},
\label{dot}
\end{equation}
in terms of the Darboux vector $\bomega=\tau{\bi t}+\kappa{\bi b}$, with dimensions of length$^{-1}$,
where the dot represents the derivative with respect to $s$.

If we take in (\ref{derivat}) the next order derivative $d^4{\bi r}/ds^4$, taking into account (\ref{dot}) and (\ref{capatau}) it can be expressed as a linear combination of the three-unit vectors where the coefficients are functions of the four derivatives. These three-unit vectors are expressed in (\ref{univect}) in terms of the derivatives of the point ${\bi r}$ up to the third order, and therefore we obtain a relationship between the derivatives of ${\bi r}$ up to order four. This means that in the Frenet-Serret formalism the most general differential equation satisfied by a continuous and differentiable geometric curve in three-dimensional space is a system of ordinary differential equations of fourth order where the independent variable is the arc length $s$.
\subsection{Physics}
Let us go one step forward. Instead of doing Geometry let us make Physics by introducing clocks. Let us consider that we describe the electron by the time evolution in three-dimensional space of a point ${\bi r}$. 
In the above geometrical analysis, it will mean that there exists a continuous and differentiable function $s(t)$ between the arc length $s$ and the elapsed time $t$, that will represent from now on the parameter of the curve ${\bi r}(t)\equiv{\bi r}(s(t))$. 
Let us consider that an inertial observer describes the free motion of this electron in terms of the time coordinate in its frame. Because in general, all derivatives are different from zero, the  motion of the point is accelerated and therefore this point ${\bi r}(t)$ cannot represent the location of the center of mass (CM) of the electron. It can represent perhaps the location of another property of the electron. 

If we are able to obtain a Lagrangian description of this system of only three degrees of freedom, the free Lagrangian $L_0$ has to depend up to the second time derivative of the point ${\bi r}$, to obtain from the Euler-Lagrange equations a system of differential equations of fourth order. If this electron interacts there will be an interaction Lagrangian $L_I$, function of the interacting parameter, the charge, the point ${\bi r}$ and some of its time derivatives and of external fields defined at point ${\bi r}$. This is telling us that the point ${\bi r}$ represents the point where the external fields are evaluated. The external force is evaluated at this point, i.e., ${\bi r}$ represents the center of charge (CC) of the electron. This point has to be a different point than the possible CM, because in the free motion the CM must describe a straight line with no acceleration.

If we are able to describe an electron by a single point ${\bi r}$, this point represents the evolution of the CC of the electron. We are not assuming any size or shape for the electron. We consider only that we have a characteristic point ${\bi r}$ that represents the center of charge where the external forces are defined, and that satisfies a system of fourth order differential equations.  

Let us consider that in an inertial reference frame the motion of point ${\bi r}$, although accelerated, represents a free motion. This means that for this inertial observer the displacement of the point at any time $t$, $ds(t)$, must be independent of the time. Otherwise, if at two different times $t_1$ and $t_2$, $ds(t_1)\neq ds(t_2)$ this means that something different is happening at time $t_2$ than at time $t_1$, that produces a different displacement, contradictory with the idea that the motion is free. For this inertial observer the free motion of the center of charge of the electron is moving at a velocity of constant absolute value. If this description has to be independent of the inertial observer, we arrive to the conclusion that we can never find an inertial observer at rest with respect to this point. 

If at some instant $t$, an instantaneous inertial observer is at rest with respect to the point ${\bi r}(t)$, the velocity of this point is 0, but because this point is accelerated, at time $t+dt$ the velocity will be different from zero, contradictory with the condition that the velocity is of constant absolute value in any inertial frame.

The general motion of the point ${\bi r}$ has curvature and torsion. If it is moving at a velocity of constant absolute value for some inertial observer, the velocity is no longer of constant absolute value for another inertial observer moving at a constant velocity ${\bi v}$ with respect to the first one, if we assume that velocities are added as in the non-relativistic formalism. If we add a constant vector ${\bi v}$ to a changing velocity vector does not produce a vector of a constant absolute value. There is nevertheless a unique solution if the velocities do not add according to the nonrelativistic linear addition.  The velocity of the CC of an electron described by a single point must necessarily be an unreachable velocity of constant absolute value for all inertial observers. The velocity of the point ${\bi r}$ must transform according to the relativistic addition formula and it must be the speed of light. This is consistent with Dirac's analysis of the electron such that the point ${\bi r}$ of Dirac's spinor $\psi(t,{\bi r})$, where the external fields are evaluated, is in fact moving at the speed of light for all inertial observers. Dirac's spinor represents the probability amplitude of the electron around the CC.

This selects that the relationship between inertial observers must contain the condition of the existence of a velocity limit for the motion of the CC of the electron, and that the relative velocity among inertial observers can never reach this limit velocity. This limit velocity is an intrinsic property and must be the same for all inertial observers. 

For the free motion also $\kappa(t)$ and $\tau(t)$, or its ratio $\kappa(t)/\tau(t)$ must be independent of the time, so that the trajectory of the CC is a helix. The center of the helix is a good candidate for the location of the CM that in the free motion is moving along a straight line at a constant velocity that can never reach the velocity of the CC. But the location of this point and the angular velocity of the comoving Frenet-Serret frame are completely determined by the point ${\bi r}$ and its subsequent time derivatives. The comoving frame rotates and the CC moves around the CM and because of this rotation and the orbital relative motion, this object has angular momentum. The CC has a {\it trembling motion} (zitterbewegung) around the CM. It is not necessary to postulate, as in De Broglie formalism, that elementary matter has an internal frequency. The particle described in this way has an internal frequency associated to this relative motion.

This simple analysis discards the use of the Galilei group as the group that defines the class of equivalent observers. If we consider that the three-dimensional space is infinite, the kinematical group of the possible formalism for describing an electron by a single point ${\bi r}$, is the Poincar\'e group and the velocity of the CC is necessarily the speed of light.

\subsection{Invariance arguments}
From another point of view, let us assume that an inertial observer describes the evolution of the electron by the trajectory of a point ${\bi r}(t)$ in three-dimensional space. If we make any arbitrary Poincar\'e transformation $g$ to any other inertial observer, what we get is the family of all trajectories ${\bi r}(t,g)$ of the point for all inertial observers, where $t$ is the time coordinate in each frame. This family is parameterized by the 10 parameters that characterize the arbitrary Poincar\'e transformation $g$. All these trajectories satisfy a Poincar\'e invariant differential equation obtained from the above family ${\bi r}(t,g)$ and its subsequent time derivatives, by eliminating the 10 parameters in terms of the different derivatives. To eliminate 10 parameters, we need to reach up to the fourth-order derivative to have sufficient equations, and therefore we find again that the trajectory of the point for any inertial observer satisfies a system of Poincar\'e invariant ordinary fourth order differential equations. This dynamical equation is Poincar\'e invariant by construction because is independent of all group parameters.

The point is not the center of mass, and in the free motion, if it represents the center of charge, it is accelerated but cannot radiate because energy is conserved. A particle with CC and CM as different points can never radiate if the CM is not accelerated. A free electron never radiates. The Lorentz-Dirac theory of radiation considers that both points are the same, but this is an unphysical constraint. The center of mass represents the location of the center of inertia of the particle while the CC represents the location of the center of interaction. Inertia and interaction are different physical properties of matter. Theoretical physics is obliged to analyze physical systems in which these two points could be different.

\section{General description}
\label{conclusions}

The previous introduction is contained basically in the Preamble of the Lecture Notes \cite{Lecture}. However, from a historical point of view, we arrived at this description of the electron by a different path. In 1985 \cite{Galileo} we analyzed the Lagrangian description of a mechanical system such that the boundary variables manifold that define the initial and final states of its variational evolution was considered to be a homogeneous space of the Galilei group. With this statement many models of non-relativistic spinning particles were described. The relativistic version of the above models was obtained in 1989 \cite{Poincare} by assuming that the boundary variables manifold of the variational formalism were homogeneous spaces of the Poincar\'e group. In that article we also assumed that the velocity of the point ${\bi r}$ could be the speed of light, as suggested by Dirac's analysis. 
Different models of relativistic spinning particles were found and also a spinning model of a classical photon described as a point and a comoving cartesian frame that rotates along the direction of the velocity. The spin of the photon has the same direction than the angular velocity, but it is an invariant mechanical property. The angular velocity and thus, the frequency of the photon, transform according to the Doppler effect transformation while the absolute value of the spin remains invariant.

This idea can be generalized to any space-time symmetry group of transformations. The kinematical group not only defines the space-time symmetries of the mechanical system. The homogeneous spaces of any Lie group are described by the same kind of parameters as the group elements, and, therefore, the set of the group parameters also supply and restrict the maximum number of classical variables to describe a classical elementary spinning particle. 

The idea underlying this definition of a classical elementary particle is based on Wigner's definition of an elementary particle in quantum formalism. According to Wigner the Hilbert space of the states of an elementary particle is a representation space of a unitary {\bf irreducible} representation of the Poincar\'e group \cite{Wigner}. Irreducible means that the complete Hilbert space can be generated from any state. By selecting any arbitrary state, we can obtain all Poincar\'e transformations of this arbitrary state. Now let us produce all finite linear combinations of these new states and finally the closure of these states by including all limits of Cauchy sequences of this kind of vectors. We have generated the whole Hilbert space. This is the meaning that the representation is irreducible. This is a group theoretical definition of elementary particle in the quantum formalism. 

The challenge is to find a definition based on group theory within the classical Lagrangian formalism. If we select any initial state of the variational description and transform this state by any Poincar\'e transformation we can obtain all possible physical states equivalent to the first one. If the mechanical system has no more classical states than the ones obtained by the above method, then, the set of states, i.e., the boundary variables manifold, is a homogeneous space of the Poincar\'e group. If the mechanical system is not an elementary particle and we consider an excited state, we can never reach that state from the ground state by a simple change of the reference frame. The set of states of a non-elementary mechanical system is not a homogeneous space of the Poincar\'e group.

Later in 1994, we showed that all Lagrangian systems obtained in this way can be quantized, producing all known relativistic and non-relativistic quantum mechanical equations. In particular, the relativistic model that the CC is moving at the speed of light is the only model that satisfies Dirac's equation when quantized \cite{Dirac}. All these results were collected in 2001 in the book \cite{Rivasbook}.

The idea that leads to the definition of a classical elementary particle was revisited in 2008 in the form of a new fundamental principle, the {\it Atomic principle} \cite{atomic}. The definition of elementary particle represents the physical fact that the elementary particle has no excited states and if it is not annihilated by some interaction its internal structure cannot be modified. A bound system like an atom can have excited states but an elementary particle cannot. Any mechanical system that contains some internal parts can be deformed but an elementary particle cannot. An elementary particle cannot be deformed and therefore all its possible states are just kinematical modifications of any one of them. In this way the whole classical theory of elementary spinning particles is based on three fundamental principles:\\
 1. The Restricted Relativity Principle, requires that the inertial observers are related by the Poincar\'e group and the dynamical equations must remain invariant under the transformations of this group. \\
 2. The Variational Principle, postulates the existence of a Lagrangian that describes the elementary particle such that the dynamical equations are the Euler-Lagrange equations, and with the above principle they must be Poincar\'e invariant. But we do not know yet what are the independent degrees of freedom. This is determined by the next fundamental principle.\\
 3. The atomic Principle, i.e., the above definition of elementary particle. If the initial state in the Lagrangian description in an inertial reference frame is $x\equiv(x_1,\ldots,x_n)$, described by an unknown set of $n$ classical variables and this state changes at a later time to another $y\equiv(y_1,\ldots,y_n)$, and if the internal structure of the elementary particle has not been modified, then it is possible to find at this later time, another inertial observer, related to the previous one by means a Poincar\'e transformation $g$, such that the variables $y$ for this observer take exactly the same values as the variables $x$ for the previous observer. Then the state $x$ is transformed into the $y$, such that the state $y=g*x$ or $x=g^{-1}*y$. And this has to be valid for any pair of states. We recover that the boundary variables manifold of the Lagrangian description is a homogenous space of the Poincar\'e group.
 
The quantum mechanical theory of elementary particles is also based on three fundamental principles. In the classical description, the Lagrangian formalism determines the path of the system between the initial point $x_1$ to the final point $x_2$. Because we are not going to measure the path followed by the system, we assume Feynman's hypothesis that all paths between these two points are allowed with the same probability. We have to substitute the Variational Principle by Feynman's Quantization Principle such that the Hilbert space that describes the quantum mechanical states is such that any wave function is a complex, squared integrable function defined on the boundary variables manifold of the corresponding classical Lagrangian description \cite{Dirac}-\cite{Rivasbook}. 

With this quantization method, the wave function $\psi(x)$ represents the probability amplitude of finding the electron around point $x$. Under an arbitrary Poincar\'e transformation $\psi(x)$ transforms with a unitary irreducible representation of the Poincar\'e group if the point $x$ belongs to a homogeneous space of the group. For other groups, like the Galilei group, what is obtained is that the wave function transforms with a {\bf projective} unitary irreducible representation of the group, i.e., a unitary representation up to an arbitrary phase.
This is a technical feature that depends on the structure of the group. If the group has nontrivial central extensions, like the Galilei group, then the representation is not a true representation but a projective representation. The Poincar\'e group has only trivial central extensions and thus all representations are true representations. In the Lagrangian formalism, Lagrangians are Poincar\'e invariant but not Galilei invariant. They transform with a time derivative of a gauge function \cite{Levy-Leblond}. 
 
The formalism can be extended to any kinematical group, Galilei, Poincar\'e, Weyl, Conformal or De Sitter groups, or any group we consider that describes the space-time symmetries of our mechanical system. If the rotation group is a subgroup of this general kinematical group, we always obtain that the elementary particles described in this formalism have spin. If for instance, we select as the boundary variables manifold the space-time, i.e. the states are defined by the variables time $t$ and the position of a point ${\bi r}$, $x\equiv(t,{\bi r})$, this manifold is in fact a homogeneous space of the Galilei or Poincar\'e group, and therefore it can be used to describe an elementary particle: The spinless point particle. The angular momentum of this particle with respect to the point ${\bi r}$ is zero. But in Nature there are no spinless elementary particles. We need to include some more variables that transform under rotations, but we are restricted to a set, larger than space-time, that belongs to a homogeneous space of the kinematical group. In that case, the angular momentum of the particle with respect to the point ${\bi r}$ is different from zero and the particle has spin.  

For the Galilei and Poincar\'e groups the maximum number of variables is 10, the number of parameters of the group. The largest homogeneous space is spanned by the variables $(t,{\bi r},{\bi u},\balpha)$, interpreted as the time $t$, the location of a point ${\bi r}$, the velocity of this point ${\bi u}=d{\bi r}/dt$ and the orientation $\balpha$, of a comoving Cartesian frame attached to the point ${\bi r}$. The three parameters $\balpha$ represent the 3 essential parameters that define a set of three orthogonal unit vectors ${\bi e}_i$, $i=1,2,3$, with origin at the point ${\bi r}$, that describe the attached Cartesian frame. The constraint $u=c$ selects, in the Poincar\'e case, the only model that satisfies Dirac's equation when quantized \cite{Dirac}. Because the Lagrangian has to depend on the next order time derivative of the above boundary variables, it is also a function of the acceleration of the point and the angular velocity of the attached comoving frame, and thus the point ${\bi r}$ satisfies a system of ordinary fourth-order differential equations. The spin observable is only related to the rotation invariance of the theory and is obtained in the Lagrangian formalism by applying Noether's theorem to the invariance under rotations. 

In the quantum version, Dirac's spinor is a squared integrable complex function of the variables $\psi(t,{\bi r},{\bi u},\balpha)$. The constraint $u=c$ implies that $\psi$ is a function of the non-compact variables $(t,{\bi r})$ and of the following five compact variables: Two are the zenithal angle $\beta$ and azimuthal angle ${\lambda}$ of the velocity ${\bi u}$ of constant value $u=c$ and the other three, the angles that describe the orientation of the attached local cartesian frame. If we represent this local frame as a rotation matrix it is characterized by the rotated angle $\alpha$ and the zenithal angle $\theta$ and azimuthal angle ${\phi}$ of the unit vector along  of the rotation axis. The spin operators are linear differential operators with respect to these 5 compact variables and commute with the Hamiltonian $H=i\hbar\partial/\partial t$ and the linear momentum operator ${\bi p}=-i\hbar\nabla$. By appropiately selecting the complete commuting set of operators, Dirac's spinor can be written is separate variables as
\[
\psi(t,{\bi r},{\bi u},\balpha)=\sum_{i=1}^{i=4} \Psi_i(t,{\bi r})\chi_i(\beta,\lambda;\alpha,\theta,\phi),
\]
where the $\Psi_i(t,{\bi r})$ describe the space-time part and the four components $\chi_i(\beta,\lambda;\alpha,\theta,\phi)$
the internal or spin part associated to the spin operators. The different representations of Dirac's equation, either Pauli-Dirac, Weyl or Majorana, etc., depend on how we describe the internal orientation operators \cite{Dirac}.
\subsection{The Classical Dirac Particle}

In addition to the previous references, we have published an article \cite{ClassicalDiracI} where we analyze in detail what we call the classical Dirac particle, i.e., the relativistic model obtained by the general formalism that satisfies Dirac's equation when quantized. All observables can be expressed in terms of the point ${\bi r}$ and its subsequent time derivatives. In particular, the center of mass (CM) ${\bi q}$ is defined with simultaneity with the CC, at every inertial frame, and it moves with a velocity ${\bi v}=d{\bi q}/dt$, such that $v<c$. The CC velocity ${\bi u}=d{\bi r}/dt$ always satisfies $u=c$. In terms of the CM velocity, it is found that the energy and linear momentum can be written as 
\begin{equation}
H=\gamma(v)mc^2,\quad {\bi p}=\gamma(v)m{\bi v},
\label{Handp}
\end{equation}
like in the case of the point particle, where the factor $\gamma(v)=(1-v^2/c^2)^{-1/2}$.
The definition at time $t$ of the CM ${\bi q}$ is
 \begin{equation}
{\bi q}={\bi r}+\frac{c^2-{\bi v}\cdot{\bi u}}{a^2}{\bi a},\quad {\bi a}=\frac{d{\bi u}}{dt},
\label{CMq}
\end{equation}
where ${\bi a}$ is the acceleration of the CC, and all magnitudes evaluated at the same time. With this definition
the fourth order differential equations of the point ${\bi r}$ can be separated into a coupled system of second order differential equations for both points ${\bi q}$ and ${\bi r}$.
 
The dynamical equations of the center of mass ${\bi q}$ and the center of charge ${\bi r}$ of the Dirac particle under any electromagnetic field obtained from the general Lagrangian $L=L_0+L_{em}$, where $L_0({\bi u},{\bi a},\bomega)$ is the free Lagrangian and $L_{em}=-eA_0(t,{\bi r})+e{\bi u}\cdot{\bi A}(t,{\bi r})$, in terms of the scalar and vector potentials $A_0$ and ${\bi A}$ defined at time $t$ at the center of charge position ${\bi r}$, respectively, are:
\begin{eqnarray}
\frac{d{\bi q}}{dt}&=&{\bi v},\quad\frac{d{\bi v}}{dt}=\frac{1}{m\gamma(v)}\left[{\bi F}-\frac{\bi v}{c^2}\left({\bi F}
\cdot{\bi v}\right)\right],\label{eq:d2qdt2}\\
\frac{d{\bi r}}{dt}&=&{\bi u},\quad \frac{d{\bi u}}{dt}=\frac{c^2-{\bi v}\cdot{\bi u}}{({\bi q}-{\bi r})^2}({\bi q}-{\bi r}),\label{eq:d2rdt2}
 \end{eqnarray}
where the external force ${\bi F}=e{\bi E}(t,{\bi r})+e{\bi u}\times{\bi B}(t,{\bi r})$, is the Lorentz force defined at the CC, and $m$ and $e$ are the mass and the electric charge of the electron, respectively. The two velocities satisfy the constraints $u=c$, $v<c$. In the derivation of these dynamical equations, we have made no use of a possible radiation reaction force. Equation (\ref{eq:d2qdt2}) is just $d{\bi p}/dt={\bi F}$, after taking the time derivative of ${\bi p}=\gamma(v)m{\bi v}$, and leaving on the left-hand side the CM acceleration $d{\bi v}/dt$. The equation (\ref{eq:d2rdt2}) is equation (\ref{CMq}) when writing on the left-hand side the CC acceleration ${\bi a}$.

The free Lagrangian $L_0({\bi u},{\bi a},\bomega)$, is translation invariant and it is an invariant function in terms of the velocity ${\bi u}$ and acceleration ${\bi a}$ of the CC, and also of the angular velocity $\bomega$ of the attached rotating local frame to the point ${\bi r}$. This local frame can be taken as the Frenet-Serret frame. The surprise is that although we have postulated the existence of a free Lagrangian, the dynamical equations and the Noether analysis of the symmetries are finally independent of this free Lagrangian because the Noether constants of the motion depend only on the partial derivatives of this free Lagrangian. It is also crucial that the point ${\bi r}$ is moving at the speed of light. All essential observables like energy, linear and angular momentum are expressed in terms of these partial derivatives. This means that any arbitrary Poincar\'e invariant function $L_0({\bi u},{\bi a},\bomega)$ of these variables will lead to the same dynamical equations and to the same definition of the fundamental observables.

\begin{figure}[!hbtp]\centering%
\includegraphics[width=5cm]{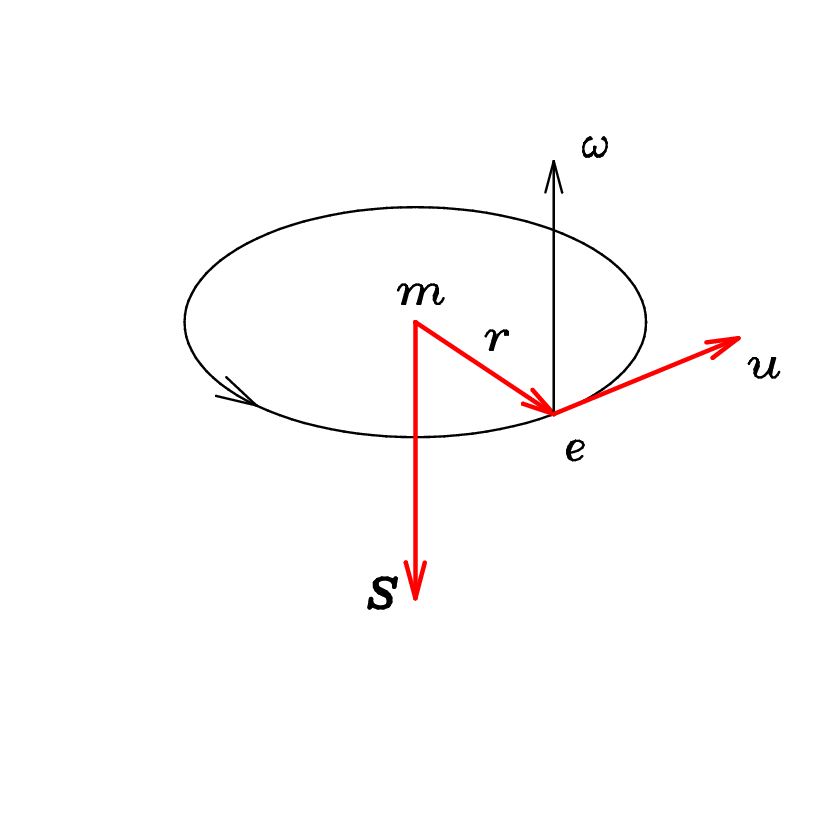}
\caption{Motion of the CC of the electron ${\bi r}$ in the center of mass frame given in (\ref{espinCM}). The CM is located at the origin in this frame. The angular velocity $\bomega$, because the CM velocity ${v=0}$, and according to (\ref{omegaunew}) has no component along the velocity ${\bi u}$, the trajectory is flat and has no torsion. The spin in the center of mass frame has the opposite direction to the angular velocity $\bomega$.} 
\label{fig1}
\end{figure}

Now the electron has two characteristic points ${\bi q}$ and ${\bi r}$ and therefore two different angular momenta can be defined with respect to both points. By the Noether analysis it is found that the angular momentum with respect to the CC ${\bi r}$, at time $t$ takes the form:
\begin{equation}
{\bi S}=\left(\frac{H-{\bi u}\cdot{\bi p}}{a^2}\right)({\bi a}\times{\bi u}),
\label{sdudtu}
\end{equation}
perpendicular to the velocity and acceleration of the CC. If we use the expressions of $H$ and ${\bi p}$ in (\ref{Handp}) and formula (\ref{CMq}) to substitute the expression of ${\bi a}$, this spin can be rewritten as
\begin{equation}
{\bi S}=-\gamma(v)m({\bi r}-{\bi q})\times{\bi u}.
\label{sCC}
\end{equation}
The angular momentum at time $t$, with respect to the CM ${\bi q}$, is defined as
\begin{equation}
{\bi S}_{CM}={\bi S}+({\bi r}-{\bi q})\times{\bi p}=-{\gamma(v)}m({\bi r}-{\bi q})\times({\bi u}-{\bi v}).
\label{sCM}
\end{equation}
These observables represent the spin of the electron ${\bi S}$ and ${\bi S}_{CM}$ with respect to the CC and CM, respectively at the same time $t$. In the center of mass frame (${\bi q}={\bi v}=0$) and both spins take the same value. They satisfy two different dynamical equations. In the free case  $d{\bi S}/dt={\bi p}\times{\bi u}$ and 
 $d{\bi S}_{CM}/dt=0$. The CM spin is conserved but the CC spin ${\bi S}$ satisfies the same dynamical equation than Dirac's spin operator in the quantum case. It corresponds to the classical spin observable equivalent to the quantum mechanical Dirac's spin operator. In the center of mass frame and for a free electron, the spin is constant and takes the value
\begin{equation}
 {\bi S}=-m{\bi r}\times{\bi u},
\label{espinCM}
\end{equation}
so that the point ${\bi r}$ describes a circle of constant radius $R_0$ at the speed of light in a plane orthogonal to the constant spin, as depicted in figure {\bf\ref{fig1}}.

In the center of mass frame, the separation $R_0=|{\bi r}-{\bi q}|$ is constant and when quantizing the model takes the value $R_0=\hbar/2mc$, and the internal motion has an angular velocity $\omega_0=2mc^2/\hbar$, twice the frequency postulated by De Broglie. The value of the spin is $S(0)=\hbar/2$.   

The angular velocity of the comoving frame $\bomega=\bomega_p+\bomega_u$, has two components, one $\bomega_p$ perpendicular to the plane subtended by ${\bi u}$ and ${\bi a}$ and that produces the curvature of the trajectory of the point and another $\bomega_u$ along the velocity ${\bi u}$, like the Darboux vector in the Frenet-Serret formalism, that gives rise to the torsion of the trajectory.
They are expressed as:
\begin{equation}
\bomega_p=\frac{1}{c^2}({\bi u}\times{\bi a}),
\label{omegaperp}
\end{equation}
\begin{equation}
\bomega_u=\left[\frac{\bomega_p\cdot{\bi v}}{c^2-{\bi v}\cdot{\bi u}}\right]{\bi u}.
\label{omegaunew}
\end{equation}

In natural units, the angular velocity is related to the Darboux vector $\bomega=\tau {\bi t}+\kappa{\bi b}$, so that the torsion is the component $\omega_u$ along the CC velocity and the curvature $\omega_p$ is along the binormal. The two components of the angular velocity of the comoving rotating frame correspond to the two intrinsic geometric properties of the trajectory of the CC in terms of time derivatives instead of derivatives with respect to the arc length $s$.

\subsection{Dirac's Hamiltonian}
There is another relation, linear in the energy and linear momentum, that depends also on the spin ${\bi S}$ and angular velocity:
\begin{equation}
H={\bi p}\cdot{\bi u}+\frac{1}{c^2}{\bi S}\cdot\left({\bi a}\times{\bi u}\right)={\bi p}\cdot{\bi u}-{\bi S}\cdot{\bomega}.
\label{eq:DiracH}
\end{equation}
The last term is always positive definite $-{\bi S}\cdot{\bomega}_p>0$, according to (\ref{sdudtu}) and (\ref{omegaperp}), because these two vectors are collinear but in opposite directions, as seen in figure {\bf\ref{fig1}}. The spin ${\bi S}$ is orthogonal to the velocity ${\bi u}$, and therefore to $\bomega_u$, and this term can be rewritten as $-{\bi S}\cdot{\bomega}>0$. The energy contains two terms: the first term, proportional to the linear momentum, represents the translational energy, while the second, proportional to the spin and angular velocity and to the motion of the CC (zitterbewegung), and independent of the CM velocity, represents the rotational energy $H_{rot}$ which is positive definite, because the spin ${\bi S}$ has the opposite direction to the angular velocity $\bomega_p$, and never vanishes. If we introduce in (\ref{eq:DiracH}) the expressions of ${\bi p}$ in (\ref{Handp}) and ${\bi S}$ in (\ref{sCC}),
\[
H=\gamma(v)m{\bi v}\cdot{\bi u}-\gamma(v)m\left[({\bi r}-{\bi q})\times{\bi u}\right]\cdot\left[\frac{1}{c^2}{\bi a}\times{\bi u}\right],
\]
we get $H=\gamma(v)mc^2$. It is this linear expression (\ref{eq:DiracH}) in terms of $H$ and ${\bi P}$, which represents the classical equivalent of Dirac's Hamiltonian as shown in \cite{Dirac}.

\subsection{Intrinsic properties of the Dirac particle}
The Classical Dirac particle has two invariant properties related to the two Casimir operators of the Poincar\'e group, namely the invariant related to the four-momentum $p^\mu\equiv(H/c,{\bi p})$, $p^\mu p_\mu=m^2c^2$, and the invariant related to the Pauli-Lubanski four-vector $w^\mu=(1/2)\epsilon^{\mu\nu\sigma\lambda}p_\nu J_{\sigma\lambda}$, that can be written in terms of $H$, ${\bi p}$ and ${\bi S}_{CM}$ as:
\begin{equation}
w^\mu\equiv({\bi p}\cdot{\bi S}_{CM},H{\bi S}_{CM}/c),\quad w^\mu w_\mu=-m^2c^2 S(0)^2,
\label{wmu}
\end{equation}
so that the two parameters, the mass $m$ and the absolute value of the spin in the center of mass frame $S(0)$, are the intrinsic properties of the electron.

Because we locate at the point ${\bi r}$ the total charge of the electron $e$, its motion with respect to the CM produces an instantaneous magnetic moment and an instantaneous electric dipole moment that are defined as \cite{magneticDirac}:
\begin{equation} 
 \bmu_{CM}=\frac{e}{m\gamma(v)}{\bi S}_{CM}=-e({\bi r}-{\bi q})\times({\bi u}-{\bi v}),
\label{eq:defMu}
\end{equation}  
\begin{equation}
{\bi d}_{CM}=e({\bi r}-{\bi q}).
\label{elecdipolMom}
\end{equation}
In this way the magnetic moment is $\mu_{CM}=e\hbar/2m$, Bohr's magneton and the electric dipole, $d_{CM}=e\hbar/2mc$. The quantization of this particle implies that the quantum electric dipole moment operator is the operator found by Dirac in 1928, but he considered no relevant from the physical point of view, perhaps for some possible loss of spherical symmetry of the charge distribution \cite{Rivasbook}. This electric dipole gives rise to Darwin's term of Dirac Hamiltonian.

All physical observables $H$, ${\bi p}$, the two spins ${\bi S}$, ${\bi S}_{CM}$, the two components of the angular velocity $\bomega_u$ and $\bomega_p$ and the synchronous location of the CM ${\bi q}$ can be expressed in terms of the point ${\bi r}$ and their time derivatives, simultaneously. Also, the electromagnetic properties magnetic moment $\bmu_{CM}$ and electric dipole moment ${\bi d}_{CM}$, if the particle has electric charge $e$.

In this analysis no hypothesis about the size or shape of the electron has been stated and no requirements concerning a possible charge or mass distribution. 
The physical properties of the electron are determined by the evolution of a single point, the {\bf center of charge}.
It is a system of only three degrees of freedom ${\bi r}$, but they satisfy a system of ordinary differential equations of fourth order and not of second order, as it corresponds to the description of the spinless point electron. In three-dimensional Euclidean space the most general continuous and differentiable curve satisfies necessarily a system of ordinary fourth order differential equations. 

The historical suggestion in some physics books that the fundamental equations of physics are second order differential equations represents a simplifying restriction. This requires that a Lagrangian description of a moving point will be a function of the point and its velocity, Euler-Lagrange equations are of second order and thus the trajectories of the point have no torsion, which is equivalent to spin suppression. It represents the mechanical description of spinless particles but in the real-world experimental physics has shown that there are no spinless elementary particles.

\section{Natural units}

The complete formalism can be done in terms of dimensionless variables once some basic properties are fixed.
For the natural system of units we take as usual $\hbar=c=1$, the mass and charge of the electron $m=e=1$, the unit of length is $2R_0=\hbar/mc=1$, and the unit of time is $\tau_0=2R_0/c=1$, where $R_0$ is the separation between the CC and the CM for the center of mass observer. Then if we define new dimensionless position variables ${\widetilde{\bi r}}={\bi r}/2R_0$, ${\widetilde{\bi q}}={\bi q}/2R_0$, new dimensionless velocity variables ${\widetilde{\bi u}}={\bi u}/c$, ${\widetilde{\bi v}}={\bi v}/c$, and a new dimensionless time ${\widetilde{t}}={t}/\tau_0$, the differential equations are written in terms of dimensionless variables, and once the tildes are removed, they become:
\begin{eqnarray}
\frac{d{\bi q}}{dt}&=&{\bi v},\quad\frac{d{\bi v}}{dt}=\frac{1}{\gamma(v)}\left[{\bi F}-{\bi v}\left({\bi F}
\cdot{\bi v}\right)\right],\label{eq:d2qdt2NU}\\
\frac{d{\bi r}}{dt}&=&{\bi u},\quad \frac{d{\bi u}}{dt}=\frac{1-{\bi v}\cdot{\bi u}}{({\bi q}-{\bi r})^2}({\bi q}-{\bi r})\label{eq:d2rdt2NU},
 \end{eqnarray}
 with the constraints $u=1$, $v<1$. In the Lorentz force ${\bi F}=e({\bi E}+{\bi u}\times{\bi B})$, if the fields are expressed in the international system of units, we need two conversion factors:
 \[
K_E=\frac{e\hbar}{m^2c^3}=7.55676\cdot10^{-19}\;{\rm m/V},
\] 
\[ 
K_B=\frac{e\hbar}{m^2c^2}=2.26546\cdot10^{-10}\;{\rm T}^{-1},
\]
such that the Lorentz force in natural units is
\[
{\bi F}=K_E{\bi E}+K_B{\bi u}\times{\bi B},
\]
with ${\bi E}$ in V/m and ${\bi B}$ in teslas. 

Therefore, this system (\ref{eq:d2qdt2NU}) and (\ref{eq:d2rdt2NU})  of ordinary non-linear differential equations, written in natural units, depend on two physical dimensionless coefficients, $K_EE$ and $K_BB$, where $E$ and $B$ are, the intensity of the electric and magnetic field, respectively, in the I.S. of units, and all remaining variables are dimensionless. The stability of this non-linear system of ordinary differential equations depends on the values of these two coefficients and therefore sets a limit on the allowed values of the external fields.

In the natural system of units, the unit of energy is $mc^2=8.18724\cdot10^{-14}$J or conversely 1 J=$1.22141\cdot10^{13}\;$natural units. The relationship between the International System of units and the Natural System of units has the following conversion factors:\\
 
\begin{tabular}{|c|c|c|}
\hline 
\rule[-1ex]{0pt}{3ex}  & {\bf IS to Natural System}& \\ 
\hline 
\rule[-1ex]{0pt}{3ex} Unit & International System & Natural System \\ 
\hline 
\rule[-1ex]{0pt}{3ex} $[L]$ length & 1 m & $2.58965\cdot 10^{12}\;$n.u. \\ 
\hline 
\rule[-1ex]{0pt}{3ex} $[M]$ mass & 1 kg & $1.09775\cdot10^{30}\;$n.u. \\ 
\hline 
\rule[-1ex]{0pt}{2.5ex} $[T]$ time & 1 s & $7.76357\cdot10^{20}\;$n.u.\\ 
\hline 
\rule[-1ex]{0pt}{2.5ex} $[Q]$ electric charge& 1 C & $6.24146\cdot10^{18}\;$n.u.\\ 
\hline 
\rule[-1ex]{0pt}{2.5ex} $[I]$ current & 1 A & $0.00804071\;$n.u.\\ 
\hline 
\rule[-1ex]{0pt}{2.5ex}  $[E]$ electric field & 1 V/m & $7.55676\cdot10^{-19}\;$n.u. \\ 
\hline 
\rule[-1ex]{0pt}{2.5ex} $[B]$ magnetic field & 1 T & $2.26546\cdot10^{-10}\;$n.u.\\ 
\hline 
\rule[-1ex]{0pt}{2.5ex} $[{\cal E}]$ energy & 1 J & $1.22141\cdot10^{13}\;$n.u. \\ 
\hline 
\rule[-1ex]{0pt}{2.5ex} $[{\cal E}]$ energy & 1 eV & $1.95693\cdot10^{-6}\;$n.u. \\ 
\hline 
\rule[-1ex]{0pt}{2.5ex} $[\epsilon_0]$ permittivity of vacuum & $1/4\pi\epsilon_0=\alpha$ & $\approx 1/137\;$ n.u.\\ 
\hline 
\end{tabular} \\

\section{Upper bound of the external fields}
The acceleration of the CM (\ref{eq:d2qdt2NU}) depends on the value of the external fields. The only constraint is that
$v<1$. Let as assume that the electron is at rest under an external electric field. In a time $\Delta t$, $\Delta v=F\Delta t=K_E E\Delta t$. The electric field is bounded by
\begin{equation}
E<\frac{1}{K_E \Delta t}=\frac{E_S}{\Delta t}.
\label{limit}
\end{equation}
If the field is acting during one unit of natural time this implies that $E<E_S$ and thus the electric field is limited above. Similarly, under an external magnetic field $B<B_S$. The above upper limits of the external fields during a unit of natural time are called the Schwinger limits
\begin{eqnarray}
E_S=\frac{1}{K_E}&=&\frac{m^2c^3}{e\hbar}=1.323\cdot10^{18}\;{\rm V/m},
\label{limitE}\\
B_S=\frac{1}{K_B}&=&\frac{m^2c^2}{e\hbar}=4.414\cdot10^9\;{\rm T}.
\label{limitB}
\end{eqnarray} 
This limitation depends on the time $\Delta t$. The unit of natural time is $\tau_0=1.288\cdot10^{-21}$s, and the time for a turn of the CC around the CM in the center of mass frame is $\pi$ in natural units. For a moving electron with a CM velocity $v$ this time for a complete cycle is $\gamma(v)\pi$.

If the transfer of energy from an electromagnetic field to matter is in the form of photons, as suggested by Planck's hypothesis, we have to compute not only the modification of the CM velocity (which represents de transfer of energy and linear momentum) but also the modification of the angular momentum, such that the electromagnetic angular momentum transfer equals to $\hbar$, the spin of the photon. This will fix the necessary time $\Delta t$ to transfer a quantum of electromagnetic field. If the discrete time $\Delta t$ to transfer a quantum of spin $\hbar$ is computed, the allowed physical electromagnetic fields are bounded above by (\ref{limit}).

\section{Mass and Spin invariance: Radiation reaction}
According to the Atomic Principle, the mass $m$ and the absolute value of the spin in the center of mass frame $S(0)$
are the invariant properties of the electron that cannot be modified by any interaction. Therefore $dm/dt=0$, and mass is conserved. Taking the time derivative of $p^\mu p_\mu=H^2-{\bi p}^2=m^2$, we get
\[
H\frac{dH}{dt}-{\bi p}\cdot\frac{d{\bi p}}{dt}=0,
\]
and if we use (\ref{Handp}) for the expressions of $H$ and ${\bi p}$ we arrive to 
\[
\frac{dH}{dt}={\bi v}\cdot{\bi F},\quad dH=d{\bi q}\cdot{\bi F}.
\]
The variation of the mechanical energy of the electron is the work of the external Lorentz force along the CM trajectory. But the work of the external field acting on a point charge is the work along the CC trajectory, $d{\bi r}\cdot{\bi F}$. If these two trajectories are different, not all electromagnetic energy expended by the field is transformed into mechanical energy of the particle. 

By energy conservation if this difference of energy ${\bi F}\cdot(d{\bi r}-d{\bi q})$, is positive the electron increases its mechanical energy, and the remaining part of this energy that has not been used in this acceleration process, must remain as energy of the field. If this difference is negative, the electron loses mechanical energy and the difference is returned and transformed into energy of the field.  It can be interpreted as if the particle radiates electromagnetic energy, because the total electromagnetic energy of the field has changed. If these two works are the same there is no radiation because all electromagnetic energy of this work is transformed into mechanical energy of the particle.

The possible radiation is related to the existence of an acceleration of the CM. The existence of an external force implies an acceleration of the CM, but it can happen that there exist CM accelerated motions without radiation.

The other invariant is that the absolute value of the spin in the center of mass frame $S(0)$ is conserved.
In the preprint \cite{spinconserv} it is analyzed this requirement of the conservation of the spin in the center of mass frame and this implies a modification of the dynamical equation (\ref{eq:d2qdt2NU}) such that the breaking term
${\bi v}({\bi v}\cdot{\bi F})$ is replaced by the term ${\bi v}({\bi u}\cdot{\bi F})$, so that this differential equation becomes
\begin{equation}
\frac{d{\bi v}}{dt}=\frac{1}{\gamma(v)}\left[{\bi F}-{\bi v}\left({\bi F}
\cdot{\bi u}\right)\right].
\label{dvdtRad}
\end{equation}
This corresponds to modify the dynamical equation $d{\bi p}/dt={\bi F}$, obtained from the Lagrangian formulation, by the equation
\begin{equation}
\frac{d{\bi p}}{dt}={\bi F}-\gamma(v)^2\left[{\bi F}\cdot({\bi u}-{\bi v})\right]{\bi v}.
\label{dpdtRad}
\end{equation}
The variation of the linear momentum contains a breaking term, opposite to the CM velocity ${\bi v}$, that depends on the factor $\gamma(v)^2$ and the difference of the work per unit time of the external Lorentz force along the CC and the CM trajectories. This term can be interpreted as the reaction force produced by the interchange of photons between the particle and the field. In a continuous interaction, the term $\gamma(v)^2\left[{\bi F}\cdot({\bi u}-{\bi v})\right]d{\bi q}$, will be the linear momentum of the continuously radiated electromagnetic field in a time $dt$, that is substracted from the impulse ${\bi F}dt$ produced by the external force.

A single electron in a finite time $T$ radiates finite electromagnetic energy. This corresponds to the emission of a finite number of photons, let us say $N$. This means that when a single electron is accelerated it is not emitting continuous radiation but rather it emits on average a single photon after a discrete time $\Delta t=T/N$.
Radiation is not a continuous process, so that we need to wait to a discrete time $\Delta t$ such that the quantum of energy corresponds to the transfer of angular momentum $\hbar$. The determination of this discrete time between two consecutive emissions of photons by a single electron in different examples, is left to future research. 

\section{Fine structure constant}
\label{sec:fine}

In (2007) we have found a Poincar\'e invariant interaction Lagrangian between two Dirac particles \cite{invariantL}.
If $x_a^\mu\equiv(ct_a,{\bi r}_a)$, $a=1,2$ are the 2 four-vectors of both particles and $\dot{x}_a^\mu\equiv(c\dot{t}_a,\dot{\bi r}_a)$, $a=1,2$,  are the $\tau$-derivatives with respect to some invariant evolution parameter $\tau$, they transform as $x_a^{'\mu}=\Lambda^\mu_\nu x_a^\nu+a^\mu$, and $\dot{x}_a^{'\mu}=\Lambda^\mu_\nu \dot{x}_a^\nu$, where $\Lambda^\mu_\nu$ is a general Lorentz transformation and $a^\mu$ a space-time translation. The following Poincar\'e invariant interaction Lagrangian
\[
\widetilde{L}_I=g\sqrt{\frac{\dot{x}_1^\mu \dot{x}_{2\mu}}{(x_1-x_2)^\mu (x_2-x_1)_\mu}}=g\sqrt{\frac{c^2\dot{t}_1\dot{t}_2-\dot{\bi r}_1\cdot\dot{\bi r}_2}{({\bi r}_1-{\bi r}_2)^2-c^2(t_1-t_2)^2}},
\]
depends on the coupling constant $g$ with dimensions of action if the evolution parameter $\tau$ is taken dimensionless. This interaction Lagrangian is also space-time scale invariant under a scale transformation that preserves the speed of light and also symmetric in the interchange $1\leftrightarrow2$. When selecting some inertial observer, the evolution parameter is taken the time coordinate $\tau=t$ in this frame, and the evolution description is a synchronous evolution, and thus $t_1=t_2=t$, $\dot{t}_a=1$, and the synchronous Lagrangian in this frame becomes
\[
L_I=g\sqrt{\frac{c^2-{\bi u}_1\cdot{\bi u}_2}{({\bi r}_1-{\bi r}_2)^2}}.
\]
This Lagrangian, without tilde in the time evolution description, has now dimensions of energy, is translation and rotation invariant but not invariant under pure Lorentz transformations, because the two particles are considered with simultaneity in this frame, and simultaneous events in one frame are not simultaneous in another moving frame.

With that Lagrangian we have analyzed numerically \cite{2elec} different electron-electron interactions. If we introduce the extra assumption that when the spin vanishes the CC and the CM become the same point, ${\bi u}_1={\bi u}_2=0$, the above interaction Lagrangian between the two Dirac particles becomes the instantaneous Coulomb interaction Lagrangian between two spinless point particles at rest, the coupling constant between the two Dirac particles is just $g=e^2/4\pi\epsilon_0 c$. This Lagrangian has dimensions of energy. The transformation coefficient from the International System of units to the Natural System for energy is the factor $1/mc^2$. The above interaction Lagrangian when written in natural units, and including the energy conversion factor, becomes:
\[
L_I=\frac{1}{mc^2}\frac{e^2}{4\pi\epsilon_0 c}\frac{c}{2R_0}\sqrt{\frac{1-{\bi u}_1\cdot{\bi u}_2}{({\bi r}_1-{\bi r}_2)^2}}=\alpha\sqrt{\frac{1-{\bi u}_1\cdot{\bi u}_2}{({\bi r}_1-{\bi r}_2)^2}}, \quad \alpha=\frac{e^2}{4\pi\epsilon_0\hbar c},
\]
$\alpha\approx1/137$, with all variables in natural units and $u_1=u_2=1$.
 
It is only through this non-relativistic limit of a Poincar\'e-invariant interaction Lagrangian and that the basic instantaneous interaction between point particles is given by Coulomb's law, that we have arrived at the conclusion that the only dimensionless parameter characterizing the interaction between two Dirac particles is the fine-structure constant. The existence of this dimensionless constant implies that the fundamental constants $\hbar$, $e$, $c$ and the permittivity of vacuum $\epsilon_0$, are not independent physical parameters.

\section{Gravitational interaction}
\label{sec:gravity}

The above non-relativistic limit of a Poincar\'e invariant Lagrangian suggests to consider whether we can arrive similarly to the non-relativistic Newtonian gravitational potential between two Dirac particles, as a consequence of their gravitational interaction.

The theory of gravitation is today Einstein's General Theory of Relativity. In the spirit of unification of all interactions, is it possible to find a Lagrangian description of the gravitational interaction between two spinning Dirac particles? In this case the coupling constant of the gravitational interaction of two Dirac particles is known as the gravitational coupling constant, and we get:
\[
\alpha_G=\frac{Gm^2}{\hbar c}=1.75175\cdot10^{-45},
\]
where $G$ is the gravitational constant and $m$ is the mass of the electron.

But this analysis of the gravitational interaction of two electrons has led us to the conclusion that today's gravitational theory is a very restrictive theory of gravitation.

Einstein's theory was formulated before the discovery in the formalism of the calculus of variations that the manifold $X$ of the boundary variables manifold of the Lagrangian description of any arbitrary mechanical system is always a metric space. It is this manifold that in this approach supplies the possible degrees of freedom and essential variables of an elementary particle.

Euler-Lagrange equations are equivalent to the geodesic equations on that manifold between the initial point $x_1\in X$ to the final point $x_2\in X$, such that the distance along the path is a minimum. 
This discovery is the result of the work of Finsler (1918) \cite{Finsler}, followed later by the paper by Cartan (1933) who called Finsler spaces to these geometric structures \cite{Cartan}. More detailed books are those of Rund (1959) \cite{Rund} and Asanov (1985) \cite{Asanov}.

The metric structure of the boundary variables manifold is characterized by a symmetric tensor $g_{ij}(x,\dot{x})$ which is a function of the point $x\in X$ but also of the derivative $\dot{x}$ with respect to some arbitrary and invariant evolution parameter $\tau$. The reason is that the Lagrangian, if the evolution is analyzed in terms of some arbitrary and invariant evolution parameter $\tau$, is a function $\widetilde{L}(x,\dot{x})$ of the boundary variables and their first order derivative with respect to $\tau$. But it is a homogeneous function of first degree in terms of the derivatives $\dot{x}_i$ of all the boundary variables.

The action integral is usually stated in terms of a time integral. But time is a relative observable and therefore the formalism expressed in terms of some particular time is not independent of the inertial observers. If we make the action integral in terms of some arbitrary invariant evolution parameter $\tau$, the same for all inertial observers, can be rewritten as:
\[
\int_{t_1}^{t_2}L\,dt=\int_{\tau_1}^{\tau_2}L\,\dot{t}(\tau)d\tau=\int_{\tau_1}^{\tau_2}\widetilde{L}\,d\tau,\quad \widetilde{L}=L\dot{t}(\tau),
\]
and $\widetilde{L}$, is a homogeneous function of first degree in the $\tau$-derivatives of the boundary variables \cite{Rivasbook}. If the evolution parameter $\tau$ is dimensionless, this $\widetilde{L}$ has dimensions of action. 

This homogeneity allows us to write the Lagrangian of any mechanical system with $n$ boundary variables $x_i$, $i=1,\ldots,n$, as a sum of as many terms as boundary variables in the form
\[
\widetilde{L}=\sum_{i=1}^{i=n}\frac{\partial \widetilde{L}}{\partial\dot{x}_i} \dot{x}_i=\sum_i F_i(x,\dot{x})\dot{x}_i,
\]
where the $F_i$ are homogeneous functions of zero degree of the $\dot{x}_i$, and thus depend on the time derivatives of the independent degrees of freedom.

The Lagrangian squared $\widetilde{L}^2$ is a homogeneous function of second degree of all derivatives $\dot{x}_i$. This allows us to write it as
\[
\widetilde{L}^2=\sum_{ij}g_{ij}(x,\dot{x})\dot{x}_i\dot{x_j}, \quad g_{ij}(x,\dot{x})=\frac{1}{2}\frac{\partial^2\widetilde{L}^2}{\partial\dot{x}_i\partial \dot{x}_j}=g_{ji},
\]
and the partial derivatives of $\widetilde{L}^2$ define the components of the symmetric metric tensor. The action integral can be rewritten as the integral of a positive function 
\[
\int_{\tau_1}^{\tau_2}\widetilde{L}(x,\dot{x})d\tau=\int_{\tau_1}^{\tau_2}\sqrt{\widetilde{L}^2(x,\dot{x})}d\tau=
\int_{\tau_1}^{\tau_2}\sqrt{g_{ij}(x,\dot{x})\dot{x}^i\dot{x}^j}d\tau=
\]
\[
=\int_{x_1}^{x_2}\sqrt{g_{ij}(x,\dot{x})d{x}^id{x}^j}=\int_{x_1}^{x_2}ds,
\]
and the action integral is the positive geodesic distance between the extremal points.

As an example, the free relativistic point particle Lagrangian
\[
L_0(t,{\bi r},{\bi u})=-mc\sqrt{c^2-{\bi u}^2},\quad \widetilde{L}_0(t,{\bi r},\dot{t},\dot{\bi r})=-mc\sqrt{c^2\dot{t}^2-\dot{\bi r}^2},
\]
$\widetilde{L}_0$ is a squared root of a second-degree polynomial in terms of all the $\dot{x}^\mu$ and therefore a homogeneous function of first degree in these derivatives. In natural units
\[
\frac{1}{2}\frac{\partial^2 \widetilde{L}_0^2}{\partial\dot{x}^\mu\partial \dot{x}^\nu}=\eta_{\mu\nu},\quad \eta_{\mu\nu}={\rm diag}(1,-1,-1,-1),
\]
where $\eta_{\mu\nu}$ is Minkowski's metric tensor. The metric of the boundary variables manifold of the free relativistic point particle is Minkowski's metric. It is a constant Finsler metric, independent of the point and of the velocity. If this particle interacts and $\widetilde{L}_I$ is the interaction Lagrangian, the new metric under interaction of the same manifold is obtained by derivation of the function $(\widetilde{L}_0+\widetilde{L}_I)^2$. But this new metric depends of the point $x^\mu$ and also on the $\dot{x}^\mu$ and it is not a Riemannian metric. Any interaction modifies the Finsler metric of the point particle.

General relativity postulates that gravity modifies the metric of the space-time or, concerning the test particle, modifies the metric of its boundary variables manifold. That postulate is unnecessary because all interactions modify the metric. The statement that the modified metric is a new tensor $g_{\mu\nu}(x)$, is a very strong restriction because the modification of the metric should include dependence also on the derivatives $g_{\mu\nu}(x,\dot{x})$. The assumption that the modified metric tensor is a pseudo-Riemannian metric is a very strong constraint because the variational formalism implies that the new modified metric is a Finsler metric. If the velocity is negligible with respect to the speed of light, implies that in the low velocity limit
\[
\lim_{v/c\to0} g_{\mu\nu}(x,\dot{x})=g_{\mu\nu}(x),
\]
produces a Riemannian metric.

We have analyzed this constraint \cite{RivasFins} and shown that General Relativity is a low velocity limit of a more general metric description of space-time. Application of General Relativity to the solar system produces reasonable results because the Sun and planets move with a small velocity compared with $c$. High velocity gravitational phenomena like very high red-shifts or gravitational collapse, which could be dependent of the velocity are left outside this possible description because Riemannian metrics are velocity independent. Other conclusions of this very restrictive theory of gravitation have to be taken with caution, including the singularities of the Schwarzschild metric.
 
The general solution of Einstein's equations can never produce a Finsler metric, because the formalism is organized to obtain as solutions only Riemannian metrics. The fundamentals of General Relativity have to be revisited to allow Finsler metric structures as solutions of the general gravitational equations.

\section{Conclusions}

In the present formalism the electron is described by the evolution of a single point ${\bi r}$, the center of charge. It is moving at the speed of light and satisfies a non-linear system of ordinary differential equations of order four. The center of mass ${\bi q}$, can be defined simultaneously in every inertial frame in terms of ${\bi r}$ and some time derivatives. With this definition the dynamical equations can be decoupled into a system of second order ordinary differential equations for both points ${\bi r}$ and ${\bi q}$. This implies that we have two different spins, the angular momentum with respect to ${\bi q}$, ${\bi S}_{CM}$, and the angular momentum with respect to ${\bi r}$, ${\bi S}$. The spin ${\bi S}_{CM}$ is conserved in the free motion but the spin ${\bi S}$ satisfies the same dynamical equation than Dirac's spin operator in the quantum case.

The existence of two different points, one the CC ${\bi r}$, associated to the electromagnetic structure, and another the CM ${\bi q}$ with the inertia of the particle, implies that when it interacts with an external electromagnetic field, there is a difference between the energy expended by the field and the mechanical energy acquired by the particle. This difference implies that the particle radiates. This modifies the dynamical equations in order that the two intrinsic properties of the particle, the mass $m$ and the absolute value of the spin $S(0)$ remain unchanged under an external interaction. The dynamical equations are modified by a radiation-reaction force, opposite to the CM velocity. 
 
Radiation is not a continuous process. Nevertheless, the classical description of the electron and Maxwell's equations are based on continuous and differentiable variables and fields. If according to Planck's hypothesis radiation is the emission of photons, these are not emitted continuously, and a discrete time must exist between two consecutive emissions of photons by any single electron, until the continuously radiated electromagnetic angular momentum amounts the quantum $\hbar$. 

The complete set of dynamical equations can be expressed in terms of dimensionless variables or natural units. In the electromagnetic interaction the only physical parameters that characterize the system of equations are the intensity of the electric and magnetic fields in natural units, so that the stability of these equations will be restricted by these values. The allowed electric and magnetic fields are not unbounded. The existence of this discrete time $\Delta t$ between the emission of consecutive photons implies that the allowed values of the external electric and magnetic fields are bounded above if the dynamical equations have to remain stable. For greater values of the fields the constraint that the CM velocity $v<1$, does not hold.

The fine structure constant is the only dimensionless parameter that characterizes a Poincar\'e invariant interaction between two Dirac particles, if the non-relativistic limit of this interaction is the Coulomb interaction.

The possible gravitational interaction between two electrons has led us to the conclusion that the foundations of General Relativity have to be revised in order to obtain Finsler metrics rather than Riemannian metrics for the description of gravitational phenomena.

In the publications and preprints \cite{ClassicalDiracI}, \cite{ClassicalDiracII}, \cite{Wolfram} we include links to some Mathematica notebooks and to Wolfram's Community Groups where we compute numerically the interaction of the Dirac particle with uniform and oscillating electric and magnetic fields, the interaction with a polarized electromagnetic plane wave and also a Poincar\'e invariant interaction of two classical Dirac particles \cite{2elec}. In this last analysis we find that two electrons with their spins parallel can form a metastable bound state which is not destroyed by electric fields and magnetic fields along the spin direction. Only a transversal magnetic field separates the two particles. This is not a Cooper pair because it is a spin 1 bound system of charge $2e$, separated by a distance below Compton's wave length, smaller than the correlation distance of the Cooper pair. 

Another conclusion of the formalism is that the so-called tunnel effect is not a pure quantum mechanical effect, because the usual analysis shows that a spinless point particle can never cross a potential barrier. But real particles are spinning particles so that the probability of crossing in quantum mechanics has to be compared with the analysis by using classical models of spinning particles. In the simple example \cite{tunnel} it is shown that even a non-relativistic spinning particle, with kinetic energy below the top of the potential, can tunnel provided its spin is orthogonal to the direction of the velocity to the barrier, while it never crosses the barrier if it is oriented in the same direction as the velocity. In any other orientation also crosses but a minimal kinetic energy below the potential barrier is required. This is known in the literature as the {\it spin polarized tunneling} effect which was used in the old magnetic hard drivers.


\end{document}